\documentclass[a4paper,11pt]{article}
\usepackage{pos}

\newcommand{\bea}{\begin{eqnarray}} \newcommand{\eea}{\end{eqnarray}}

\def\Comment#1{}

\newcommand{\bean}{\begin{eqnarray*}}
\newcommand{\eean}{\end{eqnarray*}}

\newcommand{\gapproxeq}{\lower
.7ex\hbox{$\;\stackrel{\textstyle >}{\sim}\;$}}
\newcommand{\lapproxeq}{\lower
.7ex\hbox{$\;\stackrel{\textstyle <}{\sim}\;$}}

\newcommand\lsim{\mathrel{\rlap{\lower4pt\hbox{\hskip1pt$\sim$}}
    \raise1pt\hbox{$<$}}}
\newcommand\gsim{\mathrel{\rlap{\lower4pt\hbox{\hskip1pt$\sim$}}
    \raise1pt\hbox{$>$}}}
\newcommand{\ba}{\begin{array}}
\newcommand{\ea}{\end{array}}

\newcommand{\be}{\begin{equation}}
\newcommand{\ee}{\end{equation}}
\newcommand{\bear}{\begin{eqnarray}}
\newcommand{\eear}{\end{eqnarray}}

\newcommand{\ket}{\,\rangle}
\newcommand{\bra}{\langle \,}

\newcommand{\mY}{\mathcal{Y}}
\newcommand{\Frac}[2]{\frac{\displaystyle #1}{\displaystyle #2}}

\def\bat{\begin{array}{cc}}
\title{Analysis of beyond the Standard Model resonances with effective approaches and oblique parameters}
 \ShortTitle{Analysis of BSM resonances with effective approaches and oblique parameters}

\author*[a,\dag]{Ignasi Rosell}
\notes{\note{We wish to thank the organizers for the pleasant conference. 
This work has been supported in part by the Spanish Government (PID2019-108655GB-I00, PID2020-114473GB-I00, PID2022-137003NB-I00); by the Generalitat Valenciana (PROMETEU/2021/071); by the Universidad Cardenal Herrera-CEU (INDI23/17 and GIR23/01); and by the ESI International Chair@CEU-UCH.
}}
\author[b]{Antonio Pich}
\author[c]{Juan Jos\'e Sanz-Cillero}

\affiliation[a]{Departamento de Matem\'aticas, F\'\i sica y Ciencias Tecnol\' ogicas, Universidad Cardenal Herrera-CEU, CEU Universities, 46115 Alfara del Patriarca, Val\`encia, Spain}
\affiliation[b]{IFIC, Universitat de Val\`encia -- CSIC, Apt. Correus 22085, 46071 Val\`encia, Spain}
\affiliation[c]{Departamento de F\'\i sica Te\'orica and Instituto de F\'\i sica  de  Part\'\i culas  y  del  Cosmos IPARCOS,  Universidad Complutense de Madrid, E-28040 Madrid, Spain}

\emailAdd{rosell@uchceu.es}
\emailAdd{pich@ific.uv.es}
\emailAdd{jjsanzcillero@ucm.es}

\abstract{Experiments have confirmed the presence of a mass gap between the Standard Model and potential New Physics. Consequently, the exploration of effective field theories to detect signals indicative of Physics Beyond the Standard Model is of great interest. In this study, we examine a non-linear realization of the electroweak symmetry breaking, wherein the Higgs is a singlet with independent couplings, and Standard Model fields are additionally coupled to heavy bosonic resonances. We present a next-to-leading-order determination of the oblique $S$ and $T$ parameters. Comparing our predictions with experimental values allows us to impose constraints on resonance masses, requiring them to exceed the TeV scale ($M_R \gsim\! 2\,$TeV). This finding aligns with our earlier analysis, employing a less generalized approach and the experimental bounds of that time, where we computed these observables.}

\FullConference{42nd International Conference on High Energy Physics (ICHEP2024)\\
18-24 July 2024\\
Prague, Czech Republic\\}

\begin{document}
\maketitle

\section{Introduction}

The first two runs of the Large Hadron Collider (LHC) have validated the Standard Model (SM) as the correct theory for electroweak interactions at the explored energy scales. The discovery of a Higgs-like particle~\cite{higgs}, with properties matching SM predictions, has completed the SM spectrum of fundamental fields. No new particles have been detected, suggesting a mass gap between the SM and any potential New-Physics (NP) states. This gap supports the use of effective field theories (EFTs) to systematically search for signs of heavy scales at low energies.

The construction of any EFT relies on the particle content, the symmetries, and the power counting. In the electroweak context, the power counting depends on how the Higgs field is introduced~\cite{EW1}. One can use the linear realization~\cite{EW2}, where the Higgs $h$ is part of a doublet with the three electroweak Goldstones $\vec{\varphi}$, or the more general non-linear realization~\cite{EW1}, which does not assume a specific relationship between the Higgs and the Goldstones. This work follows the non-linear approach~\cite{lagrangian}, making use of an expansion in generalized momenta. Moreover, and in addition to the non-linear electroweak EFT, which contains only SM particles, the study considers a strongly-coupled scenario with heavy bosonic resonances with $J^P=0^\pm$ and $J^P=1^\pm$ interacting with SM particles. Our previous works examined the contributions of these heavy states to the low-energy couplings of the electroweak Lagrangian~\cite{lagrangian,PRD}. 

Here we show an analysis of the constraints on the masses of heavy resonances from the electroweak oblique parameters $S$ and $T$~\cite{Peskin_Takeuchi}. Note that $S$ is associated with the NP contribution to the difference between the $Z$ self-energy at $Q^2=M_Z^2$ and $Q^2=0$, while $T$ is proportional to the NP contribution to the difference between the $W$ and $Z$ self-energies at $Q^2= 0$. Consequently, $S=T=0$ in the SM and they are defined in such a way that they are expected to be of $\mathcal{O}(1)$ in the presence of NP. 

Our earlier work of Ref.~\cite{ST} already presented a one-loop calculation of the $S$ and $T$ parameters within this strongly-coupled scenario. That analysis provided strong constraints on the Higgs couplings and heavy scales, showing that precision electroweak data require the Higgs-like scalar to have a $hWW$ coupling close to the SM prediction, and the masses of vector and axial-vector resonances to be quite degenerate and above $4\,$TeV. The larger data sets collected in recent years have allowed for more precise measurements of the $hWW$ coupling ($\kappa_W$), which is used here to update the analysis and explore a broader set of interactions, including both $P$-even and $P$-odd operators and not only the two-Goldstone ($\varphi\varphi$) and Higgs-Goldstone ($h\varphi$) cuts, but also the fermion-antifermion ($\psi \bar{\psi}$) one.

\section{The theory}

\subsection{The Lagrangian}

In the non-linear realization of Electroweak Symmetry Breaking (EWSB), operators are organized based on their behavior at low momenta, known as chiral dimensions~\cite{Weinberg}, rather than their canonical dimensions. This work focuses on the next-to-leading order (NLO) contributions to the $S$ and $T$ parameters, derived solely from the lightest absorptive cuts ($\varphi\varphi$, $h\varphi$ and $\psi \bar{\psi}$). Therefore, only operators involving at most one spin-$1$ resonance field are considered. Using the notation from Ref.~\cite{lagrangian}, the relevant $CP$-conserving Lagrangian for these calculations is
\begin{align}
\Delta \mathcal{L}_{\mathrm{RT}} &=\,\quad 
\sum_{\xi} \left[ i\,\bar\xi \gamma^\mu d_\mu \xi  
 - v \left(\bar{\xi}_L \mY \xi_R + \mbox{h.c.}\right)  
\right]  \,+\, \frac{v^2}{4}\,\left( 1 +\Frac{2\kappa_W}{v} h \right) \bra u_\mu u^\mu\ket_{2} \nonumber \\
&+\,\bra V^1_{3\,\mu\nu} \left( \Frac{F_V}{2\sqrt{2}}  f_+^{\mu\nu} + \Frac{i G_V}{2\sqrt{2}} [u^\mu, u^\nu]  + \Frac{\widetilde{F}_V }{2\sqrt{2}} f_-^{\mu\nu}  +  \Frac{ \widetilde{\lambda}_1^{hV} }{\sqrt{2}}\left[  (\partial^\mu h) u^\nu-(\partial^\nu h) u^\mu \right]   + C_{0}^{V^1_3} J_T^{\mu\nu}  \right) \ket_2 \nonumber \\
%
&+ \,\bra A^1_{3\,\mu\nu} \left(\Frac{F_A}{2\sqrt{2}}  f_-^{\mu\nu}  + \Frac{ \lambda_1^{hA} }{\sqrt{2}} \left[ (\partial^\mu h) u^\nu-(\partial^\nu h) u^\mu \right] +  \Frac{\widetilde{F}_A}{2\sqrt{2}} f_+^{\mu\nu} +  \Frac{i \widetilde{G}_A}{2\sqrt{2}} [u^{\mu}, u^{\nu}]   +  \widetilde{C}_{0}^{A^1_3}  J_{T}^{\mu\nu} \right) \ket_2 \,. 
%
%
\label{eq:Lagr}
\end{align} 
Here, $V^1_{3\,\mu\nu}$ and $A^1_{3\,\mu\nu}$ represent color-singlet custodial-triplet resonances with quantum numbers $J^{PC}=1^{--}$ ($V$) and $1^{++}$ ($A$), respectively, using an antisymmetric formalism. The Goldstone fields are parameterized through the SU(2) matrix $U = u^2 = \exp{(i \vec{\sigma}\vec{\varphi}}/v)$, where $v= (\sqrt{2} G_F)^{-1/2}=246\;\mathrm{GeV}$ is the EWSB scale. The term $u_\mu = -i u^\dagger D_\mu U u^\dagger$ involves the appropriate covariant derivative $D_\mu$  and $f_\pm^{\mu\nu}$ contains the gauge-boson field strengths. Note that couplings with a tilde are associated with odd-parity operators and we assume that they are smaller than the corresponding even-parity ones.

\subsection{High-energy constraints}

The effective resonance Lagrangian of (\ref{eq:Lagr}) contains twelve resonance parameters, including masses. To reduce the number of free couplings and, therefore, to obtain useful phenomenological bounds, we use high-energy constraints. Furthermore, this Lagrangian is an interpolation between low- and high-energy regimes, so considering a reasonable short-distance behavior is an important ingredient from a theoretical point of view.
\begin{enumerate}
\item {\bf High-Energy Vanishing of Form Factors}. The requirement that the two-Goldstone ($\pi\pi$) and Higgs-Goldstone ($h\pi$) vector and axial form factors vanish at high energies allows us to fix $G_V$, $\widetilde{G}_A$, $\lambda_1^{hA}$ and $\widetilde{\lambda}_1^{hV}$ in terms of the remaining parameters~\cite{PRD,Rosell:2023xlf,new}.
\item {\bf Weinberg Sum Rules (WSRs)}. The considered chiral symmetry of the underlying electroweak theory suggests that the $W^3 B$ correlator is an order parameter of the EWSB and it vanishes at high energies in asymptotically-free gauge theories as $1/s^3$. This leads to the 1st and 2nd Weinberg Sum Rules~\cite{WSR}: vanishing of the $1/s$ and $1/s^2$ terms, respectively. Although the 1st WSR is expected to hold in gauge theories with nontrivial ultraviolet (UV) fixed points, the applicability of the 2nd WSR depends on the nature of the UV theory. Considering both WSRs allows us to determine the combination of resonance parameters $F_V - \widetilde{F}_V$ and $F_A - \widetilde{F}_A$, which are fundamental to determine $T$ and $S$, and an additional constraint between resonance parameters~\cite{PRD,new}; if the 2nd WSR is discarded and only the 1st WSR is assumed, it is possible to determine $T$ and a lower bound of $S$~\cite{PRD,new}. 

It is important to stress that in the absence of P-odd couplings, the 1st and 2nd WSRs together require $M_A > M_V$; this mass hierarchy remains valid if odd-parity couplings are assumed to be smaller than even-parity ones, which is a reasonable assumption we take here. We will also assume $M_A>M_V$ in all dynamical scenarios, even when the 2nd WSR does not apply~\cite{new}. 

Finally, and keeping in mind that we consider that odd-parity couplings are smaller than the corresponding even-parity couplings, the 1st WSR together with the upper limit on the vector coupling $(C_0^{V^1_3})^2$ extracted~\cite{lagrangian} from LHC diboson-production studies ($W W$, $W Z$, $ZZ$, $W h$ and $Zh$; see~\cite{Dorigo:2018cbl} and references therein) allow us to demonstrate that the fermionic contributions are negligeble~\cite{new}.
\end{enumerate}



\section{Phenomenology}

In this section, we analyze the constraints on the masses of NP resonances by comparing our theoretical predictions with experimental values of the oblique parameters $S$ and $T$. The experimental values we use are from the PDG~\cite{PDG}: $S=-0.05\pm 0.07$ and $T=0.00 \pm 0.06$ with a correlation of $0.93$, and $\kappa_W=1.023\pm 0.026$. 

Let us recall the approximations made in this one-loop calculation: only the lightest absorptive cuts are considered  ($\varphi\varphi$, $h\varphi$ and $\psi \bar{\psi}$); odd-parity couplings are assumed to be subleading, leading to an expansion in $\widetilde{F}_{V,A}/F_{V,A}$; and it is considered that $M_A>M_V$ (when both WSRs are considered, this is not an assumption, but a requirement).

Taking into account what it has been explained in the previous section, two different scenarios are considered here. In the first one, both WSRs are assumed, whereas in a more conservative approach, only the 1st WSR is considered. Moreover, and although some partial results were advanced in Ref.~\cite{Rosell:2023xlf}, here we only show the main conclusions of the forthcoming work of Ref.~\cite{new}, where all the technical details will be explained in detail.

\subsection{Leading-order calculation}

\begin{figure}
\begin{center}
\includegraphics[scale=0.46]{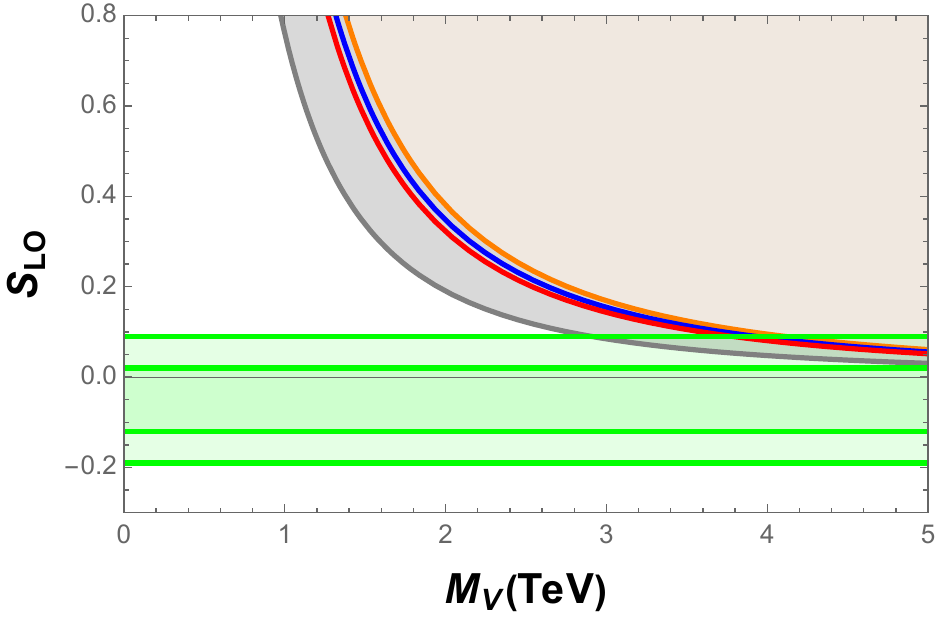}
\caption{{\small
Leading-order determination of $S$. The green area covers the experimentally allowed region, at 68\% and 95\% CL. The gray region considers the two WSRs and the corresponding lines for $M_A=M_V$ (orange), $M_A=1.1\, M_V$ (blue), $M_A=1.2\, M_V$ (red) and $M_A \to \infty$ (dark gray) are shown explicitly.  If only the 1st WSR is assumed, the allowed region is given by both, the brown and the areas areas.
}}
\label{plotSLO}
\end{center}
\end{figure}

At tree-level the $T$ parameter vanishes and we obtain the same results of Ref.~\cite{ST}, as the inclusion of P-odd operators does not affect the LO predictions. Figure 1 illustrates these predictions alongside the experimentally allowed regions at $68\%$ and $95\%$ confidence levels (CL). The gray area assumes both WSRs, with colored curves indicating specific mass ratios $M_A=M_V$ (orange), $M_A=1.1\, M_V$ (blue), $M_A=1.2\, M_V$ (red) and $M_A \to \infty$ (dark gray). When only the 1st WSR is considered, the allowed range expands to include the brown region. The experimental data suggest $M_V \!\gsim\! 2.8\,$TeV (95\% CL)~\cite{new}.

\subsection{Next-to-leading-order calculation}

\begin{figure*}
\begin{center}
\includegraphics[scale=0.46]{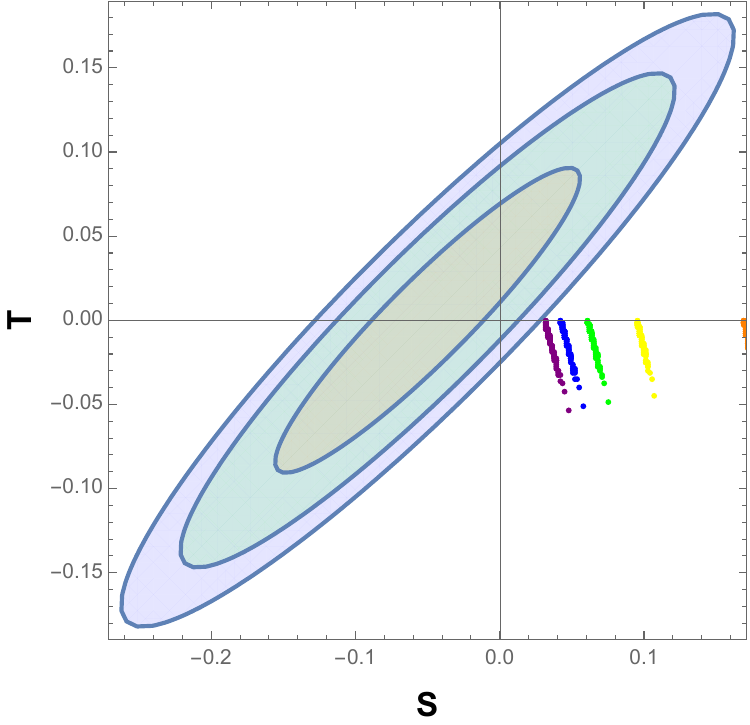} 
\includegraphics[scale=0.46]{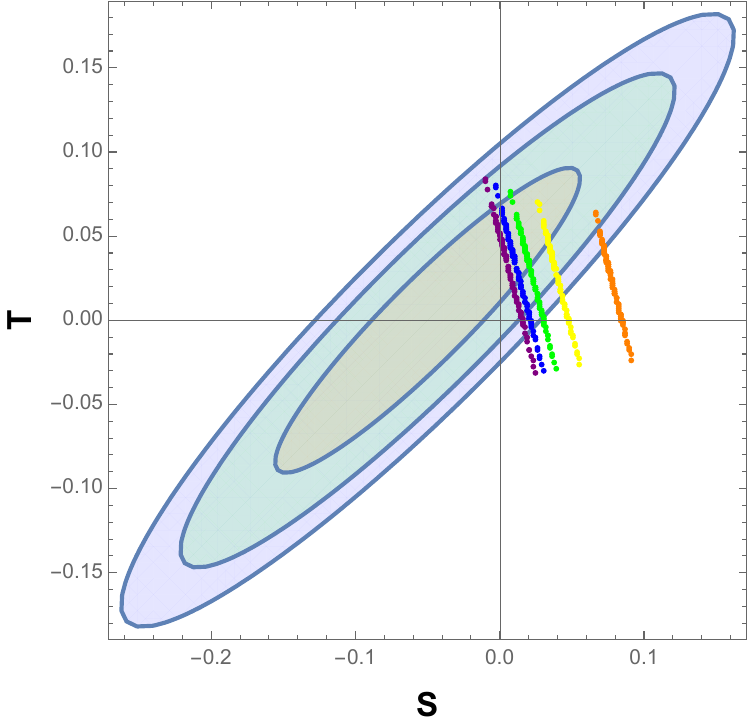} 
\includegraphics[scale=0.46]{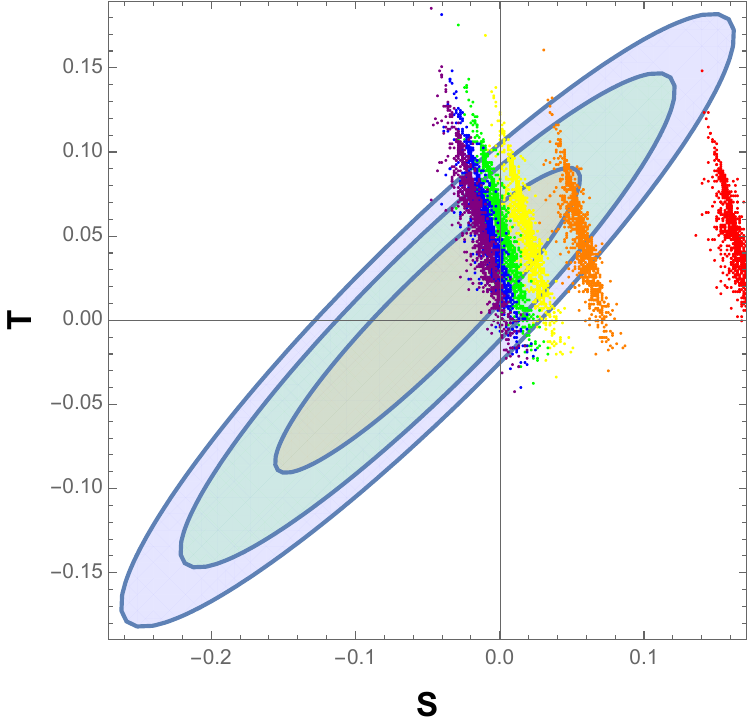} 
\includegraphics[scale=0.46]{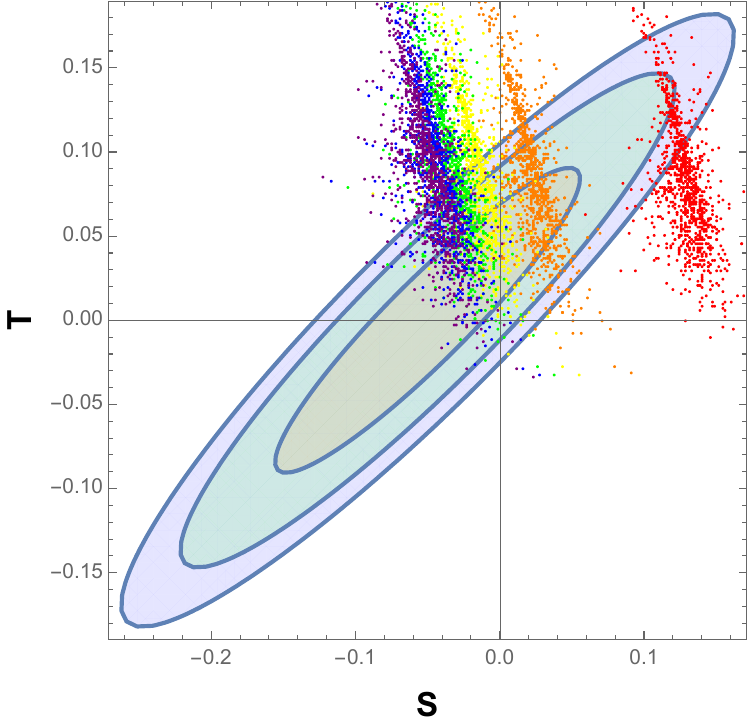}
\caption{{\small
Next-to-leading-order predictions of $S$ and $T$ considering both WSRs (top-left plot) or only the 1st WSR (top-right and bottom plots), so only lower bounds on $S$ are shown in the second case. The ellipses give the experimentally allowed regions of $S$ and $T$ at $68$\%, $95$\% and $99$\% CL~\cite{PDG}. The different colors of the points correspond to different values of $M_V$: $M_V=2$ (red), $3$ (orange), $4$ (yellow), $5$ (green), $6$ (blue) and $7$ (purple) TeV. $M_A$ is fixed in terms of the other resonance parameters if the 2nd WSR is assumed and we have considered different values in terms of $M_V$ when the 2nd WSR is discarded: $M_A=M_V$ (top-right), $M_A=1.2\; M_V$ (bottom-left) and $M_A=1.5\; M_V$ (bottom-right). The values of $\kappa_W$ and $\widetilde{F}_{V,A}/F_{V,A}$   have been obtained by considering normal distributions given by $\kappa_W=1.023\pm 0.026$~\cite{PDG} and $\widetilde{F}_{V,A}/F_{V,A}=0.00\pm 0.33$. }}
\label{fig:NLO_1WSR}
\end{center}
\end{figure*}

Once all the short-distance constraints of Section 2.2 have been implemented, the NLO determinations of $S$ and $T$ are given in terms of only four free parameters: $M_V$, $M_A$, $\widetilde{F}_{V}/F_{V}$ and $\widetilde{F}_{A}/F_{A}$. Moreover, the last two parameters are expected to be small, and we assume a normal distribution $\widetilde{F}_{V,A}/F_{V,A}=0.00\pm 0.33$. 

When one considers both WSRs, there is an additional constraint allowing to fix $M_A$ close to $M_V$ and the predictions are shown in the top-left plot of Figure 2, where different values of $M_V$ have been taken into consideration: $M_V=2$ (red), $3$ (orange), $4$ (yellow), $5$ (green), $6$ (blue) and $7$ (purple) TeV. Note that the experimental data imply $M_V \gsim\! 7\,$TeV~\cite{new}. 

If one considers the more general scenarios where only the 1st WSR is fulfilled, $M_V$ and $M_A$ are independent and in Figure 2 we consider the same values of $M_V$ quoted previously and for $M_A$ we show the cases $M_A=M_V$ (top-right), $M_A=1.2\; M_V$ (bottom-left) and $M_A=1.5\; M_V$ (bottom-right). Note that the discard of the 2nd WSR enlarges the possible values of $M_V$ to $M_V \gsim\! 2\,$TeV~\cite{new}.


\end{document}